\documentclass[twocolumn]{aastex631}

\usepackage{natbib}
\usepackage{lineno}
\usepackage{comment}

\begin{document}

\title{Quasar Negative Feedback to Surrounding Galaxies Probed with Ly$\alpha$ Emitters and Continuum-Selected Galaxies}

\author[0009-0000-2525-9236]{Yuta Suzuki}
\affiliation{Graduate School of Science and Engineering, Ehime University 2-5 Bunkyo-cho, Matsuyama, Ehime 790-8577, Japan; yuta@cosmos.phys.sci.ehime-u.ac.jp}

\author[0000-0001-5063-0340]{Yoshiki Matsuoka}
\affiliation{Research Center for Space and Cosmic Evolution, Ehime University 2-5 Bunkyo-cho, Matsuyama, Ehime 790-8577, Japan}

\author[0000-0003-3214-9128]{Satoshi Kikuta}
\affiliation{Department of Astronomy, School of Science, The University of Tokyo, 7-3-1 Hongo, Bunkyo-ku, Tokyo 113-0033, Japan}

\author[0000-0002-0673-0632]{Hisakazu Uchiyama}
\affiliation{Department of Advanced Sciences, Faculty of Science and Engineering, Hosei University, 3-7-2 Kajino-cho, Koganei, Tokyo 184-8584, Japan}
\affiliation{National Astronomical Observatory of Japan, 2-21-1, Osawa, Mitaka, Tokyo 181-8588, Japan}

\author[0000-0002-3801-434X]{Haruka Kusakabe}
\affiliation{Department of General Systems Studies, Graduate School of Arts and Sciences, The University of Tokyo
3-8-1 Komaba, Meguro-ku, Tokyo, 153-8902, Japan}
\affiliation{National Astronomical Observatory of Japan, 2-21-1, Osawa, Mitaka, Tokyo 181-8588, Japan}

\author[0000-0001-6186-8792]{Masatoshi Imanishi}
\affiliation{National Astronomical Observatory of Japan, 2-21-1 Osawa, Mitaka, Tokyo 181-8588, Japan}
\affiliation{Department of Astronomy, School of Science, Graduate University for Advanced Studies (SOKENDAI), Mitaka, Tokyo 181-8588, Japan}
\affiliation{Toyo University, 5-28-20, Hakusan, Bunkyo-ku, Tokyo 112-8606, Japan}

\begin{abstract}
We report on the statistical analysis of quasar photoevaporation at $z\sim2.2$ by comparing the density of surrounding Ly$\alpha$ Emitters (LAEs) and continuum-selected galaxies, based on the imaging data of Hyper Suprime-Cam (HSC) Subaru Strategic Program (SSP) and CFHT Large Area $U$-band Deep Survey (CLAUDS).
We select 18 quasars from Sloan Digital Sky Survey (SDSS) in the HSC Deep/UltraDeep fields, normalize the LAE/continuum-selected galaxy distribution around each quasar with the quasar proximity size, stack them, and then measure the average densities of the galaxies.
As a result, we find that the density of LAEs is $\gtrsim 5 \sigma$ lower than that of continuum-selected galaxies within the quasar proximity region.
Within the quasar proximity region, we find that the LAEs with high Ly$\alpha$ equivalent widths (EWs) are less dense than those with low EWs at the 3$\sigma$ level and that LAEs with EW of $\gtrsim150\,\mathrm{\AA}$ (rest-frame) are predominantly scarce. 
Finally, we find that both LAEs and continuum-selected galaxies have smaller densities when they are closer to quasars.
We argue that the photoevaporation effect is more effective for smaller dark matter haloes predominantly hosting LAEs, but that it may also affect larger haloes.
\end{abstract}

\section{Introduction} \label{sec:intro}
Quasars are one of the most energetic objects in the Universe and emit strong ultraviolet (UV) radiation.
In vicinity of quasars, the local UV radiation ionizes and heats surrounding neutral gas. 
The ionized regions propagate between galaxies and eventually reach nearby dark matter (DM) haloes.
The neutral gas in the DM halo is dispersed by ionization and heating, and galaxy formation is prevented because cooling and gravitational collapse do not proceed.
This phenomenon is called ``photoevaporation''.
In simulations, the photoevaporation effect may heavily suppress galaxy formation, in paticular at the low-mass end \citep[e.g.,][]{Kitayama2000, Kitayama2001, Benson2002}.
This negative feedback is often cited to explain why high redshift quasars are not always found in overdensities, because the local UV radiation around quasars is much stronger than UV background radiation.

In observational studies, two type of galaxies, Lyman Break Galaxies (LBGs) and Ly$\alpha$ Emitters (LAEs), are the promised tools for identifying high-$z$ quasar environments.
LAEs are usually selected by using a narrowband (NB) filter to capture Ly$\alpha$ emission line, while LBGs are selected by combining broadband (BB) filters to capture a Lyman break in the continuum.
LAEs are less massive and are younger than LBGs in general \citep[e.g.,][]{Ono2010, Firestone2025}.
Deep, wide-field imaging observations are required to measure the quasar environment, and Suprime-Cam \citep[S-Cam;][]{Miyazaki2002} on the Subaru Telescope had been powerful in this field.
Several studies reported a possible presence of the photoevaporation effect.
\citet{Kashikawa2007} observed LBGs and LAEs around a quasar at $z=4.87$.
They found that LBGs are distributed in a filamentary structure involving the quasar, whereas LAEs are ring-shaped in distribution avoiding the quasar vicinity within a distance of $\sim0.77\,\mathrm{pMpc}$.
\citet{Utsumi2010} found that LBGs cluster around a quasar at $z=6.43$ but appear to avoid within 2 pMpc from the quasar, and later \citet{Goto2017} found the lack of LAEs within 10 pMpc from the same quasar.
\citet{Ota2018} investigated a quasar field at $z=6.61$, and reported that LAEs show underdensity while LBGs show overdensity within 3 pMpc from the quasar.
These studies concluded that the lack of galaxies around quasars is due to the suppressed formation of galaxies by the photoevaporation.
On the other hand, other studies did not find evidence for photoevaporation effect.
\citet{Kikuta2017} examined distribution of LAEs/LBGs around two quasars at $z\sim4.9$, and found no signature of the quasar photoevaporation for LAEs at least down to $L_{\mathrm{Ly\alpha}}\sim10^{41.8}\,\mathrm{erg\,s^{-1}}$.
\citet{Bosman2020} studied a quasar field at $z=5.79$ and argued that three LAEs in the quasar proximity region were not affected by photoevaporation.
Thus, the picture of photoevaporation effect by quasars is still controversial, which is likely at least due to small sample sizes and the different conditions of past observations.
\citet{Uchiyama2019} carried out observations of LAEs in 11 quasar fields at $z=2-3$.
They found that LAEs with high Ly$\alpha$ $\mathrm{EW_{0}}$ are relatively scarce in the vicinity of quasars, supporting the quasar photoevaporation.
However, scarcity of LAEs may also be explained by a scenario in which quasar environments predominantly contain a more evolved population, such as LBGs.
In order to clarify which scenario is supported, we need simultaneous sampling of low-mass (LAEs) and high-mass (LBGs) galaxies.

Hyper Suprime-Cam \citep[HSC;][]{Miyazaki2018, Komiyama2018}, the successor to S-Cam, is more powerful to investigate quasar environments.
HSC Subaru Strategic Program (HSC-SSP) survey provides imaging data in three layers, i.e., Wide, Deep, and UltraDeep, with varying combination of the area and depth of observations \citep{Aihara2018a, Aihara2018b, Aihara2019, Aihara2022}.
Especially in Deep/UltraDeep (D/UD) layers, the survey used not only five BB filters \citep[$grizy$;][]{Kawanomoto2018}, but also NB filters, namely, NB387, NB816, NB912, and NB1010.
Based on the imaging data obtained with these NB filters, LAE catalogs have been created and used in a program entitled {\it Systematic Identification of LAEs for Visible Exploration and Reionization Research Using Subaru HSC} ({\it SILVERRUSH}) \citep[e.g.,][]{Ono2021, Kikuta2023}.
The most recent catalog in \citet{Kikuta2023} makes use of the data set combining the HSC-SSP BB/NB images from the Public DR 3 and that from the Subaru intensive program named Cosmic HydrOgen Reionization Unveiled with Subaru \citep[CHORUS;][]{Inoue2020}.
The catalog contains an unprecedentedly large sample of LAE candidates at $z=$2.2, 3.3, 4.9, 5.7, 6.6, 7.0, and 7.3.
In addition the Canada-France-Hawaii Telescope (CFHT) Large Area $U$-band Deep Survey \citep[CLAUDS;][]{Sawicki2019} covers a large portion of D/UD layers, and provides $U$-band imaging data of depth comparable to HSC-SSP images.
We can improve photometric redshifts in particular at $z=2-3.5$ by complementing $grizy$ with $U$-band photometry, which captures Balmer/4000\,\AA\,breaks and Lyman break \citep{Sawicki2019}.
\citet{Desprez2023} utilize HSC-SSP+CLAUDS data to create a continuum-selected galaxy catalog containing $U$+$grizy$ photometry and photometric redshifts.

This work presents an analysis of quasar environments by using both LAEs and continuum-selected galaxies.
LAEs represent low-mass galaxies, while continuum-selected galaxies represent high-mass galaxies like LBGs.
Using the galaxy samples taken from the HSC-based catalogs, we investigate 18 quasar fields at $z=2.2$ in order to study the photoevaporation effect on low- and high-mass galaxies.

This letter is organised as follows: 
Section \ref{sec:data} describes the data and sample selection of quasars, LAEs, and continuum-selected galaxies.
In Section \ref{sec:results}, we present a methology to measure the density of LAEs and continuum-selected galaxies around quasars.
We discuss the possible implication of the results in Section \ref{sec:discussion}.
Throughout this letter, we assume a Lambda cold dark matter ($\mathrm{\Lambda}$CDM) cosmology with $\Omega_{\mathrm{m}}=0.3$, $\Omega_{\mathrm{\Lambda}}=0.7$, $H_{0}=70\,\mathrm{km\,s^{-1}\,Mpc^{-1}}$ which are consistent with the values reported by \citet{Planck2020}.
We use the AB magnitude system \citep{Oke1983}.

\section{Data and sample selection} \label{sec:data}
\subsection{LAEs} \label{subsec:lae}
We use the {\it SILVERRUSH} LAE catalog of \citet{Kikuta2023}.
Their catalog includes 6,995 LAEs at $z=2.2$ selected with the NB387 filter, in the D/UD layers covering a total of 23.94 deg$^{2}$.
The NB387 filter is centered at $3863\,\mathrm{\AA}$ with the full width at half maximum (FWHM) of $55\,\mathrm{\AA}$, which covers the Ly$\alpha$ emission at $z_{\mathrm{Ly\alpha}}=2.178\pm0.023$.
In this sample, rest-frame Ly$\alpha$ equivalent widths (EW$_{0}$) are $\gtrsim15$ $\mathrm{\AA}$.

\begin{figure*}[t!]
\begin{center}
\hspace*{-2.5cm}
\includegraphics[scale=0.18]{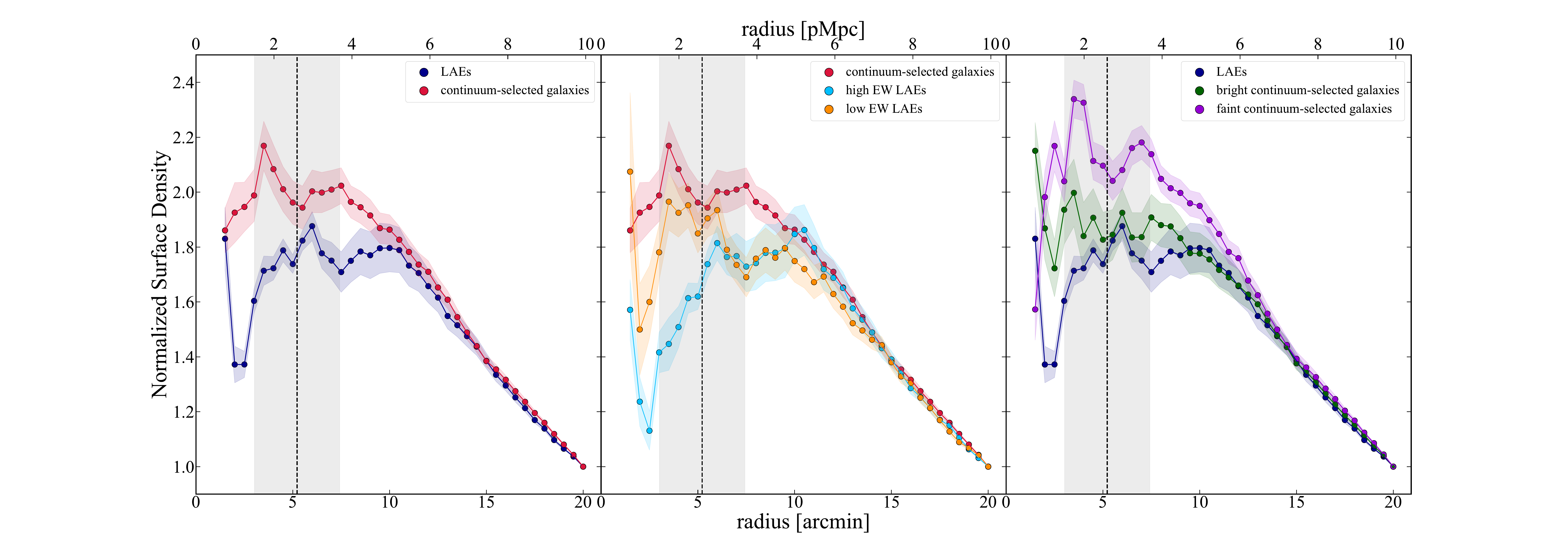}
\end{center}
\vspace*{-0.5cm}
\caption{
Radial number density profile of galaxies measured by stacking the 18 quasar fields, normalized at 20 arcmin.
The color shaded region shows the 1$\sigma$ error calcurated by bootstrap resampling method.
The dashed line and grey shaded region show the median proximity radius and its 1$\sigma$ error of the present quasar sample.
{\it Left}: Comparison of LAEs (blue) and continuum-selected galaxies (red).
{\it Middle}: Comparison of the continuum-selected galaxies, high EW LAEs (light blue) and low EW LAEs (orange).
{\it Right}: Comparison of the LAEs, bright continuum-selected galaxies (green) and faint continuum-selected galaxies (purple).
}
\label{radialprofile}
\end{figure*}

\subsection{continuum-selected galaxies} \label{subsec:gal}
We use the catalog of \citet{Desprez2023}.
\citet{Desprez2023} used two sets of codes to build the catalogs.
One is \texttt{SExtractor} and \texttt{Le Phare}.
The other is the HSC-SSP photometric pipeline \texttt{hscPipe} and \texttt{Phosphoros}.
We use the former sets, which were reported to perform slightly better in terms of photometric redshifts.
It has been shown that the brighter galaxies, the better the photometric redshift.
In this catalog, a good agreement is observed between the photometric and spectroscopic redshifts at magnitudes $i<25.0$, with a scatter $\sigma \lesssim 0.04$ and a small fraction of outliers $\eta \lesssim 7 \%$ with $|z_{\mathrm{spec}}-z_{\mathrm{photo}}| \le 0.15(1+z_{\mathrm{spec}})$ \citep[see detail in][]{Desprez2023}.
In this letter, we selected galaxies with cmodel $i<25.0$ and tighter redshift range $|z_{\mathrm{quasar}}-z_{\mathrm{photo}}| \le 0.05(1+z_{\mathrm{quasar}})$, where $z_{\mathrm{quasar}}$ is the spectroscopic redshift of a quasar.
We removed galaxies with $z_{\mathrm{photo\_err}} > 0.1$.
Almost all of our continuum-selected galaxies are labeled as star-forming galaxies in the catalog (3861 star-forming galaxies and 39 passive galaxies).

\subsection{quasars} \label{subsec:qso}
Our quasar sample is extracted from the Sloan Digital Sky Survey (SDSS) DR16 quasar caltalog \citep{Lyke2020}.
We selected 18 quasars in the D/UD fields whose redshift corresponds to $z_{\mathrm{Ly\alpha}}=2.178\pm0.023$.
The quasar UV luminosity range is $\log\lambda L_{\mathrm{\lambda}}(912\,\mathrm{\AA}) / [\mathrm{erg \, s^{-1}}]=45.0-46.8$.
We retrieved spectral data of the quasars from the SDSS Science Archive Server (SAS), in order to estimate the quasar proximity region size.

\section{Analysis and Results} \label{sec:results}
We estimate quasar proximity region size as follows.
The proximity region is the region in which the UV radiation from a quasar is higher than the UV background radiation.
The quasar radiation feedback, including photoevaporation effect, is expected to be most prominent in the quasar proximity region. 
In terms of UV intensity, we have $J_{\mathrm{\nu}}=F_{\mathrm{\nu}}/4 \pi$ where $J_{\mathrm{\nu}}$ is in units of erg s$^{-1}$ cm$^{-2}$ Hz$^{-1}$ str$^{-1}$.
Using the UV intensity at the Lyman limit $J_{21}$, we have the following expression for a UV intensity $J_{\mathrm{\nu}}$ \citep{Kashikawa2007, Uchiyama2019}:
%
%
\begin{equation}
J_{\mathrm{\nu}} = J_{\mathrm{21}}\biggl(\frac{\nu}{\nu_{\mathrm{L}}}\biggr)^{\alpha} \times 10^{-21} \, \mathrm{erg\,cm^{-2}\,s^{-1}\,Hz^{-1}\,sr^{-1}},
\end{equation}
%
%
where $\nu_{\mathrm{L}}$ is the frequency at the Lyman limit, and $\alpha$ is the continuum slope.
Since $J_{\mathrm{\nu}}=L_{\mathrm{\nu}}/(4 \pi r)^{2}$, where $L_{\mathrm{\nu}}$ is the quasar luminosity and $r$ is the distance to the quasar, the radius $r_{\mathrm{prox}}$ at which the local UV radiation is equal to the UV background radiation is given by
%
%
\begin{equation}
r_{\mathrm{prox}}=\frac{1}{4\pi} \sqrt{\frac{L_{\mathrm{\nu_{L}}}}{J_{\mathrm{21}}^{\mathrm{bkg}}(\frac{\nu}{\nu_{\mathrm{L}}})^{\alpha} \times 10^{-21}}},
\end{equation}
where $L_{\mathrm{\nu_{L}}}$ is a luminosity at the Lyman limit.
The intensity of the UV background radiation at the Lyman limit has been evaluated to be $J_{\mathrm{21}}^{\mathrm{bkg}}=1.0^{+0.5}_{-0.3}$ at $z=2$ from the quasar proximity effect measurements \citep{Cooke1997}.
We estimate quasar $L_{\mathrm{\nu_{L}}}$ by fitting a single power law to the quasar spectra at $1340-1360 \, \mathrm{\AA}$, $1440-1450 \, \mathrm{\AA}$, and $1700-1730 \, \mathrm{\AA}$, which are free from strong emission lines.

We normalize (streched or shrinked) the spatial scale of each quasar field, in such a way that the proximity region size becomes 5.2 arcmin ($\sim$ 2.5 pMpc at $z\sim2.2$), which is the median of the whole sample.
We then stack all the normalized images, and counted the number of LAEs or continuum-selected galaxies as a function of distance from the quasar. 

We compare the density of LAEs/continuum-selected galaxies in Figure \ref{radialprofile}.
The densities are normalized at 20 arcmin.
The two samples have comparable density at $>10$ arcmin when normalized.
The LAEs density profile becomes flat at the smaller distance, while that of the continuum-selected galaxies continue to increase toward the quasars.
The difference between the two samples is most prominent within the proximity region at $\lesssim5$ arcmin, where the LAE density is lower at a level of $>5\sigma$ significance.
Next, we devide the LAE sample into ``high EW'' and ``low EW'' subsamples at the median rest-frame equivalent width of EW$_{0} = 75 \, \mathrm{\AA}$ (middle panel).
We find that the density of high EW LAEs is $\sim3\sigma$ lower on average than that of low EW LAEs in the quasar proximity region.
We also devide the continuum-selected galaxy sample into ``faint'' and ``bright'' subsamples at the rest-frame UV absolute magnitude $M_{\mathrm{UV}}=-20.7$ (right panel, see Sect.4 for more detailed discussion).

\begin{figure}[t!]
\begin{center}
\vspace*{-1.0cm}
\hspace*{-0.5cm}
\includegraphics[scale=0.3]{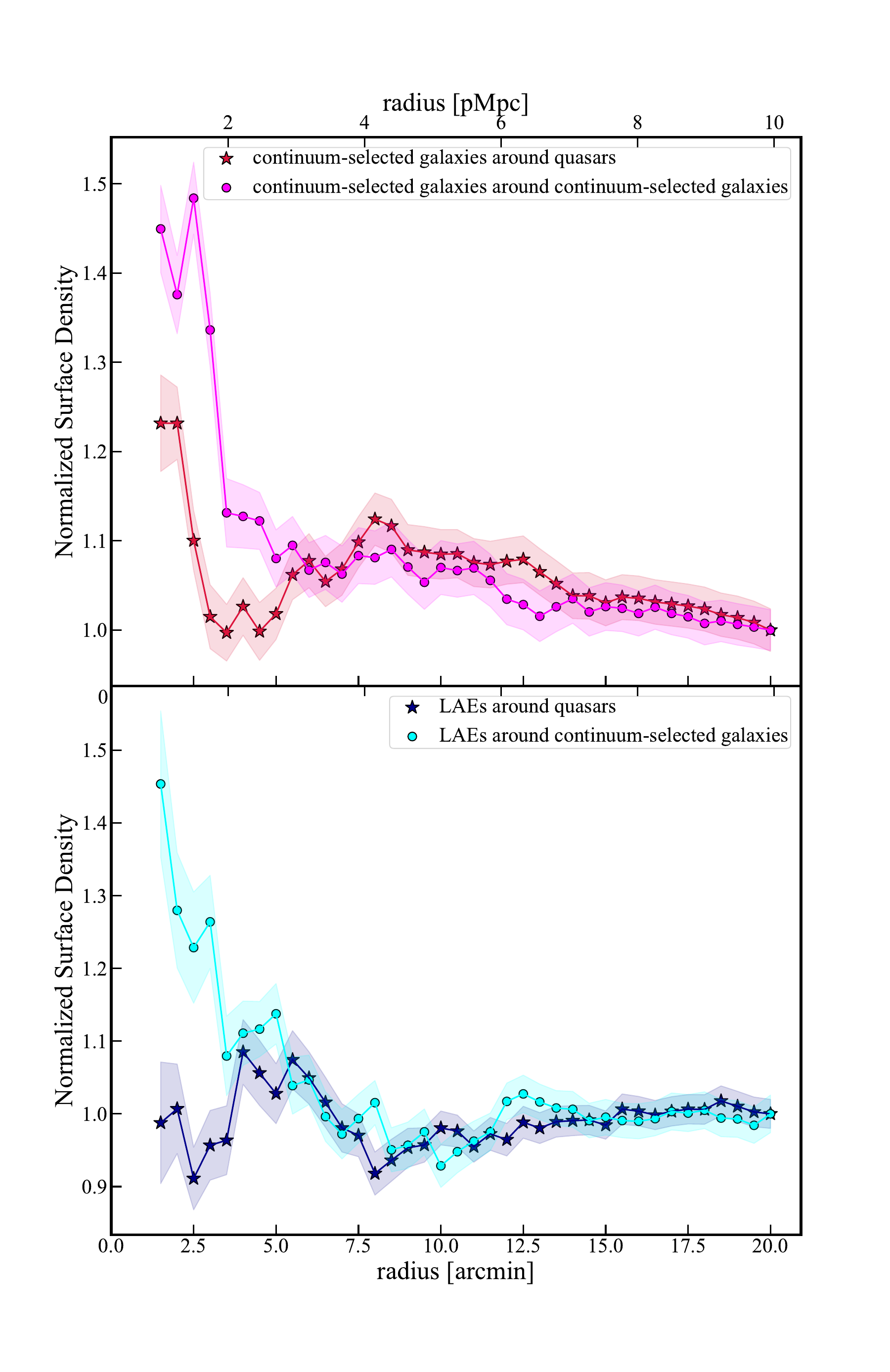}
\end{center}
\vspace*{-1.25cm}
\caption{
Radial number density profile measured by stacking the 18 quasar fields and continuum-selected galaxy fields, normalized at 20 arcmin.
The halo masses of continuum-selected galaxies are comparable to that of quasars.
The shaded region shows the $1\sigma$ uncerainties calcurated by the bootstrap resampling method.
(Top) Comparison of the density of continuum-selected galaxy around the control continuum-selected galaxies (magenta circle) and around quasars (red star).
(Bottom) Comparison of the density of LAE around the control continuum-selected galaxies (cyan circle) and around quasars (blue star).
}
\label{randomradial}
\end{figure}

Finally, we compare the quasar environments with blank fields.
\citet{Eftekharzadeh2015} reported the average halo mass of quasars to be $2.72^{+0.28}_{-0.27} \times 10^{12} \, h^{-1} \, M_{\odot}$ at $\bar{z}=2.29$, for slightly higher luminosities ($\log L=45.6-47.5$ erg s$^{-1}$) compared to the present sample. 
For continuum-selected galaxies, \citet{Harikane2022} estimated a halo mass of star-forming galaxies at $z\sim2.2$ using the HSC-SSP+CLAUDS data.
The luminosity range of their sample is essentially the same as that of our continuum-selected galaxy sample.
There is a relation between halo mass and UV luminosity of galaxies expressed in the following double power-law function:
$M_{\mathrm{h}} = \frac{M_{\mathrm{h, 0}}}{2}[10^{-0.4(M_{\mathrm{UV}}-M_{\mathrm{UV, 0}})\alpha} + 10 ^{-0.4(M_{\mathrm{UV}}-M_{\mathrm{UV, 0}})\beta}]$,
where $\log M_{\mathrm{h, \, 0}}=11.82$, $M_{\mathrm{UV, \,0}}=-19.77$, $\alpha=0.61$, and $\beta=2.34$ \citep{Harikane2022}.
Assuming this correlation, we randomly select continuum-selected galaxies with halo mass comparable to that of quasars.
In this comparison, we do not perform the normalization of the spatial scales both for the quasar sample and the galaxy sample.
Figure \ref{randomradial} compares the galaxy density around continuum-selected galaxies and around quasars.
We find that both LAEs and continuum-selected galaxies are clustered, with stronger clustering around the continuum-selected galaxies than around the quasars.

\section{Discussion and Summary} \label{sec:discussion}
We found that the normalized density of LAEs is $\gtrsim5\sigma$ lower than that of continuum-selected galaxies in the quasar proximity region (Figure \ref{radialprofile}).
Futhermore, the density of high EW LAEs is $\sim3\sigma$ lower than that of low EW LAEs.
We argue below that these results can be explained by the photoevaporation effect.

\begin{figure}[t!]
\begin{center}
\vspace*{-1.0cm}
\hspace*{-0.5cm}
\includegraphics[scale=0.45]{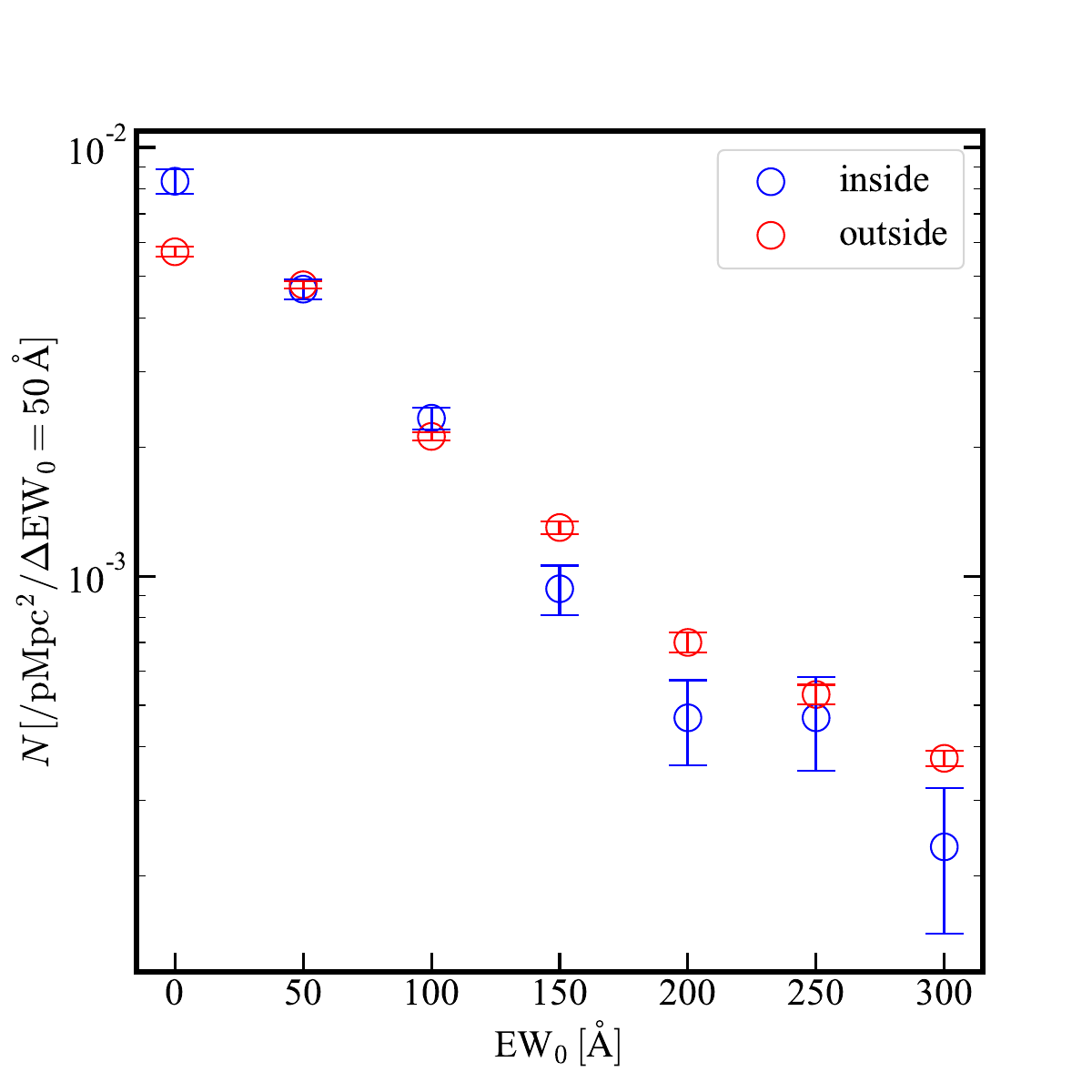}
\end{center}
\vspace*{-0.6cm}
\caption{
Distribution of EW$_{0}$ of LAEs measured by stacking the 18 quasar fields.
The blue/red circle show the distribution inside/outside the quasar proximity region.
The error bar shows the $1\sigma$ uncerainties calcurated by the bootstrap resampling method.
}
\label{EW}
\end{figure}

Figure \ref{EW} compares the EW distribution of LAEs inside and outside of the proximity region. 
At EW$_{0}$ $\gtrsim 150 \mathrm{\AA}$, the LAE densities inside the proximity zone is $\sim 2 \sigma$ lower than those outside.
It is well known that EW$_{0}$ of Ly$\alpha$ emission and stellar mass of LAEs have anti-correlation \citep[e.g.,][]{Nilsson2009, McCarron2022}.
\citet{Goovaerts2024} found a linear relationship in the form of $\log(M_{\ast}/M_{\odot}) = (-0.62 \pm 0.07) \log(\mathrm{EW_{0}}) + (9.1 \pm 0.1)$.
Futhermore, \citet{Kusakabe2018} measured the stellar-to-halo-mass-ratio (SHMR) of LAEs at $z\sim2.2$ to be $0.02^{+0.07}_{-0.01}$, with a slightly fainter limiting magnitude than our LAEs sample (NB387 $<$ 25.5 mag).
Using these relations, we can estimate the halo mass $M_{\mathrm{h}} = 2.8^{+9.9}_{-1.4} \times 10^{9} \, M_{\odot}$ for LAEs with $\mathrm{EW_{0}} = 150 \mathrm{\AA}$.
\citet{Kashikawa2007} estimated a delay time of star formation due to local UV radiation as a function of intensity $J_{21}$ and virial halo mass $M_{\mathrm{vir}}$, based on the hydrodynamical model of \citet{Kitayama2000, Kitayama2001}.
They suggested that the delay amounts to $\gtrsim20$ Myr for with $M_{\mathrm{vir}} \lesssim 3 \times 10^{9} \, M_{\odot}$ and $J_{21}\gtrsim1$.
Assuming that quasars have comparable lifetime of $\sim30$ Myr \citep[e.g.][]{Martini2004,Hopkins2006, Shen2007}, we can argue that LAEs with $\mathrm{EW_{0}} \gtrsim 150 \mathrm{\AA}$ cannot exist around a quasar while it is active.
These discussions are consistent with \citet{Uchiyama2019}.


Another possible scenario is that galaxies have already evolved into high-mass galaxies in the vicinity of quasars, resulting in the reduced density of high-EW LAEs \citep{Uchiyama2019}.
The host haloes collapsed and underwent initial starburst before quasar active phase. 
The galaxies in those haloes has already finished LAE phase and are detected as continuum-selected galaxies.
Figure \ref{EW} shows that the density of LAEs with EW$_{0}=0-50$ $\mathrm{\AA}$ inside of the proximity region is significantly higher than that of outside of the proximity region.
This indicates that vicinity of quasars may be slightly clustered and mature, and massive LAEs are increasing in the vicinity of quasars.
However, this scenario would contradict the trend in Figure \ref{randomradial}, where both LAEs and continuum-selected galaxies are more clustered around continuum-selected galaxies than around quasars.
Since the quasars and the control continuum-selected galaxies have the same halo masses, the difference in the clustering amplitudes must have something to do with a quasar activity.
The trend in Figure \ref{randomradial} may also be due to the photoevaporation effect, for continuum-selected galaxies.
\citet{Bruns2012} showed that intense UV radiation may suppress the star formation in massive haloes up to $M_{\mathrm{vir}} = 1.2 \times 10^{12} \, M_{\odot}$ by a semi-analytic model.
Adopting their model, we can explain the reduced density of continuum-selected galaxies around quasars as the effect of photoevaporation.
We also found a similar result that the number density of $U$-dropout galaxies is significantly lower around SDSS quasars than around $U$-dropout galaxies in our previous study \citep{Suzuki2024}.

\begin{figure}[t!]
\begin{center}
\hspace*{-0.5cm}
\includegraphics[scale=0.45]{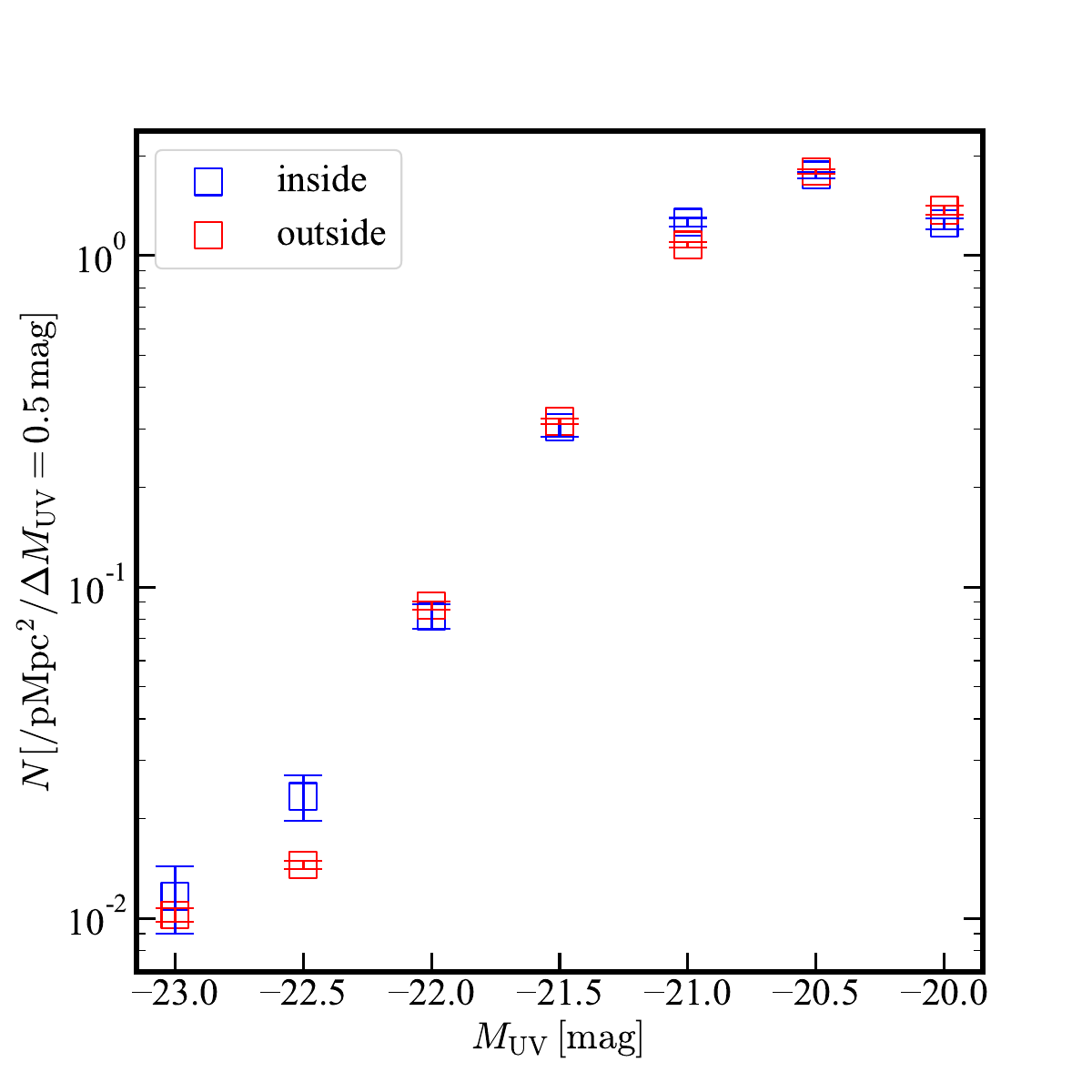}
\end{center}
\vspace*{-0.6cm}
\caption{
Distribution of $M_{\mathrm{UV}}$ of continuum-selected galaxies measured by stacking the 18 quasar fields.
The blue/red square show the distribution inside/outside the quasar proximity region.
The error bar shows the $1\sigma$ uncerainties calcurated by the bootstrap resampling method.
}
\label{UV}
\end{figure}

Figure \ref{UV} shows that the density of continuum-selected galaxies.
The decrease in density at $M_{\mathrm{UV}}=-20.0$ mag is due to an effect of incompleteness.
We found in Figure \ref{radialprofile} that the density of faint continuum-selected galaxies, which are supposed to be less massive, is higher than that of bright continuum-selected galaxies.
We also investigate the UV absolute magnitude distribution of continuum-selected galaxies inside and outside of the proximity region (Figure \ref{UV}), and the number of inside continuum-selected galaxies is consistent within 1$\sigma$ of the error margin.
This discrepancy may be due to contamination, as it reduces the accuracy of phorometric redshift measurements in the faint continuum-selected galaxies.
It is possible that continuum-selected galaxies classified as $z=2.2$ actually have different redshifts and increasing in density.
This issue will be addressed by spectrographic observations such as PFS.
This discussion may also be influnced by the difference in the probed volume between LAEs ($\Delta z \sim0.05$; 65 Mpc) and continuum-selected galaxies ($\Delta z \sim 0.15$; 200 Mpc). 
However, in the quasar proximity region, the density of LAEs is flat and that of high EW LAEs shows a decline (Figure \ref{radialprofile}).
This supports the idea that quasar radiation may have suppressed the formation of low-mass LAEs \citep{Uchiyama2019}.


%
%
%

\vskip \baselineskip
\section*{Acknowledgments}
Y.S. was supported by JST, the establishment of university fellowships towards the creation of science technology innovation, Grant Number JPMJFS2131 and by JST SPRING, Japan Grant Number JPMJSP2162.
Y.M. was supported by the Japan Society for the Promotion of Science (JSPS) KAKENHI grant No. 21H04494.
S.K. was supported by the Japan Society for the Promotion of Science (JSPS) KAKENHI grant No. 24KJ0058 and No. 24K17101.
H.U. is grateful for the support provided by JSPS KAKENHI grant (24K00684, 22K14075).
H.K. was supported by the Japan Society for the Promotion of Science (JSPS) KAKENHI grant No. 23KJ2148 and No. 25KJ17444.

The Hyper Suprime-Cam (HSC) collaboration includes the astronomical communities of Japan and Taiwan, and Princeton University.
The HSC instrumentation and software were developed by the National Astronomical Observatory of Japan (NAOJ), the Kavli Institute for the Physics and Mathematics of the Universe (Kavli IPMU), the University of Tokyo, the High Energy Accelerator Research Organization (KEK), the Academia Sinica Institute for Astronomy and Astrophysics in Taiwan (ASIAA), and Princeton University.
Funding was contributed by the FIRST program from Japanese Cabinet Office, the Ministry of Education, Culture, Sports, Science and Technology (MEXT), the Japan Society for the Promotion of Science (JSPS), Japan Science and Technology Agency (JST), the Toray Science Foundation, NAOJ, Kavli IPMU, KEK, ASIAA, and Princeton University. 
 This paper makes use of software developed for the Large Synoptic Survey Telescope.
We thank the LSST Project for making their code available as free software at  http://dm.lsst.org 
The Pan-STARRS1 Surveys (PS1) have been made possible through contributions of the Institute for Astronomy, the University of Hawaii, the Pan-STARRS Project Office, the Max-Planck Society and its participating institutes, the Max Planck Institute for Astronomy, Heidelberg and the Max Planck Institute for Extraterrestrial Physics, Garching, The Johns Hopkins University, Durham University, the University of Edinburgh, Queen's University Belfast, the Harvard-Smithsonian Center for Astrophysics, the Las Cumbres Observatory Global Telescope Network Incorporated, the National Central University of Taiwan, the Space Telescope Science Institute, the National Aeronautics and Space Administration under Grant No. NNX08AR22G issued through the Planetary Science Division of the NASA Science Mission Directorate, the National Science Foundation under Grant No. AST-1238877, the University of Maryland, and Eotvos Lorand University (ELTE) and the Los Alamos National Laboratory. 
Based on data collected at the Subaru Telescope and retrieved from the HSC data archive system, which is operated by Subaru Telescope and Astronomy Data Center at National Astronomical Observatory of Japan.
This work is based on data collected at the Subaru Telescope and retrieved from the HSC data archive system, which is operated by the Subaru Telescope and Astronomy Data Center at the National Astronomical Observatory of Japan.

These data were obtained and processed as part of the CFHT Large Area U-band Deep Survey (CLAUDS), which is a collaboration between astronomers from Canada, France, and China described in \citet{Sawicki2019}.
CLAUDS is based on observations obtained with MegaPrime/ MegaCam, a joint project of CFHT and CEA/DAPNIA, at the CFHT which is operated by the National Research Council (NRC) of Canada, the Institut National des Science de l'Univers of the Centre National de la Recherche Scientifique (CNRS) of France, and the University of Hawaii.
CLAUDS uses data obtained in part through the Telescope Access Program (TAP), which has been funded by the National Astronomical Observatories, Chinese Academy of Sciences, and the Special Fund for Astronomy from the Ministry of Finance of China.
CLAUDS uses data products from TERAPIX and the Canadian Astronomy Data Centre (CADC) and was carried out using resources from Compute Canada and Canadian Advanced Network For Astrophysical Research (CANFAR).

\bibliographystyle{aasjournal}
\bibliography{reference}

\end{document}